\documentclass[draft,jgrga]{agu2001}
\usepackage{graphicx}

\authorrunninghead{A. L. MISHEV, I. G. USOSKIN, G.A. KOVALTSOV}

\titlerunninghead{NEUTRON MONITOR YIELD FUNCTION}

\journalid{XXX 2004} \articleid{XXX}{AX} \paperid{xxxxxxxxxx}
\cpright{AGU}{2004}

\received{Month XX, 2010} \revised{Month XX, 2010} \accepted{Month XX, 2010} \published{}

\authoraddr{Corresponding author: I. Usoskin,
Sodankyl\"a Geophysical Observatory (Oulu unit), University of Oulu, Oulu, Finland
(Ilya.Usoskin@oulu.fi)}

\begin{document}

\title{Neutron Monitor Yield Function: New Improved computations}

 \authors{A. L. Mishev,\altaffilmark{1,3}
 I. G. Usoskin,\altaffilmark{1,2}
 G.A. Kovaltsov, \altaffilmark{4}}

\altaffiltext{1}{Sodankyl\"a Geophysical Observatory (Oulu unit), University of Oulu, Oulu, Finland.}

\altaffiltext{2}{Department of Physics, University of Oulu, FIN-90014, Finland.}

\altaffiltext{3}{Institute for Nuclear Research and Nuclear Energy, Bulgarian Academy of Sciences, Sofia, Bulgaria.}

\altaffiltext{4}{Ioffe Physical-Technical Institute of Russian Academy of Sciences, St. Petersburg, Russia.}

\begin{abstract}
A ground-based neutron monitor is a standard tool to measure cosmic ray variability near Earth, and
 it is crucially important to know its yield function for primary cosmic rays.
Although there are several earlier theoretically calculated yield functions, none of them agrees
 with experimental data of latitude surveys of sea-level neutron monitors, thus suggesting for an inconsistency.
A newly computed yield function of the standard sea-level 6NM64 neutron monitor is presented here
 separately for primary cosmic ray protons and $\alpha-$particles, the latter representing
 also heavier species of cosmic rays.
The computations have been done using the GEANT-4 Planetocosmics Monte-Carlo tool and a realistic
 curved atmospheric model.
For the first time, an effect of the geometrical correction of the neutron monitor effective area, related to the
 finite lateral expansion of the cosmic ray induced atmospheric cascade, is considered, that was neglected in
 the previous studies.
This correction slightly enhances the relative impact of higher-energy cosmic rays (energy above 5--10 GeV/nucleon)
 in neutron monitor count rate.
The new computation finally resolves the long-standing problem of disagreement between the theoretically calculated
 spatial variability of cosmic rays over the globe and experimental latitude surveys.
The newly calculated yield function, corrected for this geometrical factor, appears fully consistent with the
 experimental latitude surveys of neutron monitors performed during three consecutive solar minima
 in 1976--77, 1986--87 and 1996--97.
Thus, we provide a new yield function of the standard sea-level neutron monitor 6NM64 that
 is validated against experimental data.
\end{abstract}

\begin{article}


\section{Introduction}

The Earth is constantly bombarded by energetic particles - cosmic rays (CRs), which consist of nuclei of various elements, mostly protons and $\alpha-$particles.
The lower energy part (below several tens of GeV/nucleon) of the galactic cosmic ray (GCR) energy spectrum is  modulated
 in the solar wind and the heliospheric magnetic field which are ultimately related to solar activity \citep[e.g.,][]{potgieter98}.
The main instrument to study the cosmic ray variability is the world network of neutron monitors \citep{shea_SSR_00,moraal00}.
A neutron monitor (NM) \citep{simpson58} is an instrument that provides continuous recording of the hadron component of atmospheric secondary particles
 \citep{hatton71}, that is related to the intensity of high-energy primary nuclei impinging on the Earth's atmosphere from space.
The purpose of the neutron monitor is to detect variations of intensity in the interplanetary cosmic ray spectrum.
A major challenge is the reconstruction of primary particle characteristics from ground-based data records, since the
 NM is an energy-integrating device measuring only the secondaries (mainly protons and neutrons) deep in the atmosphere.
The task of linking NM count rates to the intensity of primary CRs is non-trivial and requires extensive numerical simulations of the
 atmospheric cascade, initiated by energetic CR \citep[e.g.,][]{clem00}.
A standard way to do it is to calculate the NM yield function, viz. response of a standard NM to the unit intensity of
 primary CR particles with fixed energy.

Several attempts have been performed to calculate the NM yield function.
Because of the complexity of atmospheric cascades, Monte-Carlo simulation provides the most
 useful tool to calculate the yield function.
\citet{debrunner82} used a state-of-the-art specifically developed Monte-Carlo simulation of the atmospheric cascade \citep{debrunner68}.
\citet{clem00} applied the FLUKA Monte-Carlo package for the simulations.
\citet{matthiaPhD}, \citet{matthia09} and \citet{flueckiger08} used the GEANT-4 Monte-Carlo package.
Because of the different model assumptions (hadron interaction models, atmospheric models, numerical schemes)
  the results are somewhat different (within 10-15\%) between different packages \citep{heck06,bazilevskaya08}.
While a NM is a standard device, the local surrounding may vary quite a bit for individual NMs (efficiency of the detection system,
 and local environment \citep[e.g.,][]{kruger08}), which may cause deviation of $\pm 15$\%
 from the "standard" unit \citep{usoskin_Phi_05}.
Therefore, it is hardly possible to directly calibrate the yield function for a given NM and define which one is more correct.

However, there is an indirect way to test the NM yield function, via latitude surveys \citep{clem00,caballero12}.
During a sea-level latitude survey, a NM is placed onboard a ship which scans all geomagnetic latitudes, from the equator to
 the polar region.
Accordingly, the measured count rate can be calculated via the geomagnetic shielding and the yield function of a NM
 and assuming the constant primary CR intensity during the survey.
For the latter, surveys are typically done during periods of solar minima.
Moreover, the local environment remains constant during the cruise.
A number of NM latitude surveys have been performed in the past \citep{potgieter79,moraal89,villoresi00} to provide
 a basis to validate the NM yield function.
\citet{caballero12} have made a detailed analysis of experimental data and demonstrated that none of the previously theoretically
 calculated yield functions can reproduce the observed latitude surveys.
This was ascribed primarily to the potential role of obliquely incident primary CR particles \citep{clem00}.
All the theoretically calculated earlier yield functions predict too strong latitudinal dependence of the NM
 count rate compared to the observed one, suggesting that the contribution of higher energy CRs is somewhat underestimated in the models
 compared to the lower energy part.

In order to resolve the situation, \citet{caballero12} presented an empirically derived NM yield function, which
 is the derivative of an NM latitude survey de-convoluted with the prescribed energy spectrum of GCR.
This empirical approach, while fixing the discrepancy related to latitude surveys,
 has however its own weak point -- the high rigidity (above 16--17 GV,
 which is the vertical geomagnetic cutoff rigidity at the magnetic equator) tail of the thus defined yield function is based
 solely on an extrapolation of the latitude survey,
Thus, this empirically-based ad-hoc yield function remains unclear for the energy range above this rigidity/energy range,
 which is responsible for half or more of the NM count rate \citep{ahluwalia10}.
Moreover, this method is slightly model-dependent and cannot distinguish yields for protons and $\alpha-$particles.
Thus, there is still a crucial need for a theoretical updated computation of the NM yield function that would
 agree with the experimental data.

Here we present a new, improved computation of the yield function of a standard 6NM64 neutron monitor.
In our model we considered an effect neglected in previous theoretical computations of the yield function --
 increase of the effective area of an NM as function of the energy of primary CR due to lateral extent of the
 atmospheric cascade for high energy particles.
This effect is responsible for higher, than earlier though, response of a NM to high energy CRs and, as a consequence,
 leads to perfect agreement with the measured latitude surveys.

In Section~\ref{Sec:YF} we describe the detail of the NM yield function computation.
The newly computed NM yield function is presented in Table~\ref{Tab:YF} and Fig.~\ref{Fig:YF_p},
 and confronted  with the measured latitude surveys in Section~\ref{Sec:survey}.
Conclusions are given in Section~\ref{Sec:conc}.

\section{Computation of the NM yield function}
\label{Sec:YF}
The total response of a NM to CRs can be determined by convolution of the cosmic ray spectra with the yield function.
The count rate of a NM at time $t$ is presented as:
\begin{equation}
N(P_{c},h,t)=\sum_{i}\int_{P_{c}}^{\infty}Y_{i}(P,h)\ J_{i}(P,t)\ dP
\label{Eq:N}
   \end{equation}
 where $P_{c}$ is the local geomagnetic cutoff rigidity \citep{smart_geo_06}, $h$ is the
 atmospheric depth (or altitude).
The term $Y_{i}(P,h)$ [m$^2$ sr] represents the NM yield function for primaries of particle type $i$,
 $J_{i}(P,t)$ [GV m$^2$ sr sec]$^{-1}$ represents the rigidity spectrum of primary particle of type $i$ at time $t$.
The NM yield function is defined as \citep[e.g.][]{flueckiger08}
\begin{equation}
Y_{i}(P,h)= \sum_{j} \int \int A_{i}(E,\theta)\cdot F_{i,j}(P,h,E,\theta)\ dE\ d\Omega
\label{Eq:Y}
   \end{equation}
where $A_{i}(E,\theta)$ is the geometrical detector area times the registration efficiency,
   $F_{i,j}$ is the differential flux of secondary particles of type $j$ (neutrons, protons, muons, pions) for
   the primary particle of type $i$, $E$ is the secondary particle's energy, $\theta$
   is the angle of incidence of secondaries.
Thus, the yield function consists of two parts.
One is the cascade in the Earth's atmosphere that produces secondary particles, viz. the term $F$ in Eq.~\ref{Eq:Y},
 and the other is the response of the detector itself, the $A$-term, to these secondary particles, mainly neutrons and  protons.
We will consider these two parts separately.

As the standard NM we consider here a sea-level 6NM64 \citep{stoker00,moraal00}.

\subsection{Cosmic ray induced atmospheric cascade}

As the first step, we obtained the atmospheric flux of various secondary CR particles, as denoted by term $F$ in Eq.~\ref{Eq:Y}.
The computations were performed on the basis of simulations with PLANETOCOSMICS code \citep{desorgher05},
 based on GEANT-4 \citep{agostinelli03}, similar to \citep{flueckiger08,matthia09}.
The main contribution to the NM count rate is due to secondary neutrons and protons \citep{hatton71, clem00, dorman04}.
As an input for the simulations, we used primary particles (protons and $\alpha$-particles) with the fixed energy ranging from
 100 MeV/nucleon to 1 TeV/nucleon that impinge the top of the atmosphere with a given angle of incidence.
A realistic curved atmospheric model NRLMSISE2000 \citep{picone02} was employed for the simulations.
The computations were normalized per one primary particle.
We recorded secondary particles (protons and neutrons separately) with energy within the energy bin $E_n\pm\Delta E$
 that cross a given horizontal level at the atmospheric depth 1033 g/cm$^2$, corresponding to the sea-level.
The result of simulations corresponds to the flux of secondary neutrons and protons with given energy across a horizontal unit area.

We explicitly consider cosmic ray particles heavier than protons, primarily $\alpha-$particles.
After the first inelastic collision in the atmosphere, the primary nucleus is disintegrated and is effectively
 represented by nucleons, so that from the point of view of the atmospheric cascade at the sea level, e.g. an
 $\alpha$-particle is almost equivalent to four protons with the same energy per nucleon \citep{engel92}.
However, because of the lower $Z/A$ ratio, heavier species are less modulated by both geomagnetic field and in the heliosphere.
Therefore, heavier species should be considered explicitly, not through scaling protons.
On the other hand, $\alpha-$particles are representative for all GCR species heavier than proton \citep{webber03,usoskin_bazi_11},
 as their rigidity-to-energy per nucleon ratio, and thus the modulation/shielding effect, is nearly the same,
 as well as their cascade initiation ability expressed per nucleon with the same energy.
The assumption that all the heavier species can be effectively represented by $\alpha-$particles via simple scaling
 might lead to an additional uncertainty in the middle-upper atmosphere above the height of about 15 km, but it works
 well in the lower atmosphere.
It has been numerically tested that nitrogen, oxygen and iron are very similar to $\alpha$-particle
 (scaled by the nucleonic number with the same energy per nucleon) in the sense of the atmospheric cascade at
 the sea-level \citep[e.g.][]{usoskin_COST_08,mishev11}.

Since the contribution of obliquely incident primaries is important \citep{clem97}, in particular for the
 analysis of solar energetic particle events \citep{cramp97,vashenyuk06}, we simulated cascades induced by protons with various angles of incidence
 on the top of the atmosphere.

The difference in the fluxes of secondary particles at the sea level between vertical and 15$^\circ$ inclined primary protons is not significant,
 but the secondary particle flux decreases with further increase of the incident angle, as the mass overburden increases, resulting on NM yield function decrease.
In the forthcoming results we consider isotropic (within $2\pi$ steradian) flux of incoming primary GCR.

As many as $10^6$ cascades were simulated for each energy point and for each type of primary particle (proton or $\alpha$).
The atmospheric cascade simulation is explicitly performed with PLANETOCOSMICS code for primary $\alpha$-particles in the
 energy range below 10 GeV/nucleon using the same assumptions and models as for protons.
For the energy above 10 GeV/nucleon we substitute an $\alpha$-particle with four nucleons (protons), similarly to recent works
 \citep{usoskin_JGR_06,mishev11}.

\subsection{NM registration efficiency}
\label{Sec:geom}

As the detector's own registration efficiency, denoted as term $A$ in Eq.~\ref{Eq:Y},
 we apply the results by \citep{clem00} in an updated version [J. Clem, private communications, 2011--2012].

\begin{figure}
\begin{center}
\resizebox{8cm}{!}{\includegraphics{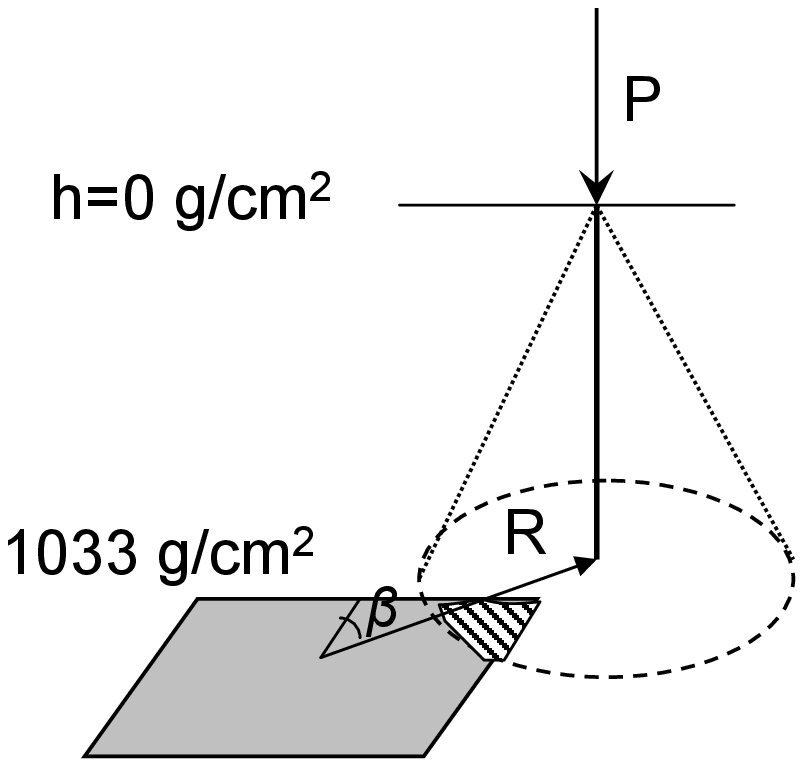}}
\end{center}
\caption{
A scheme illustrating the way to calculate the geometrical factor $G$ of the NM effective size.
The gray shape represents the physical size of the NM at the sea level, while the dashed line corresponds to the lateral extend of the
 hadronic component of an atmospheric cascade started on the top of the atmosphere.
The hatched area denotes that secondaries of the cascade hit the detector.
The scheme illustrates a general situation when a cascade, initiated by a primary CR particle with rigidity $P$ impinging on the
 top of the atmosphere, has its axis at distance $R$ and azimuth angle $\beta$ from the NM center.
\label{Fig:S}}
\end{figure}

We noted that there is process, which has not been accounted for in previous studies, that increases the
 effective area of a NM for more energetic cosmic rays, compared to the physical area of the detector.
This is the finite lateral extension of the atmospheric cascades, which is typically neglected when considering the NM yield function.
It is normally assumed that only cascades with axes hitting the physical detector's body cause response.
However, even cascades which are outside the detector may contribute to the count rate, as illustrated in Fig.~\ref{Fig:S}.
The size of a 6NM64 monitor is about $2\times 6$ m, and a typical size of the secondary particle flux spatial span
 (full width at half magnitude) at the sea level is several meters.

\begin{figure}
\begin{center}
\resizebox{\columnwidth}{!}{\includegraphics{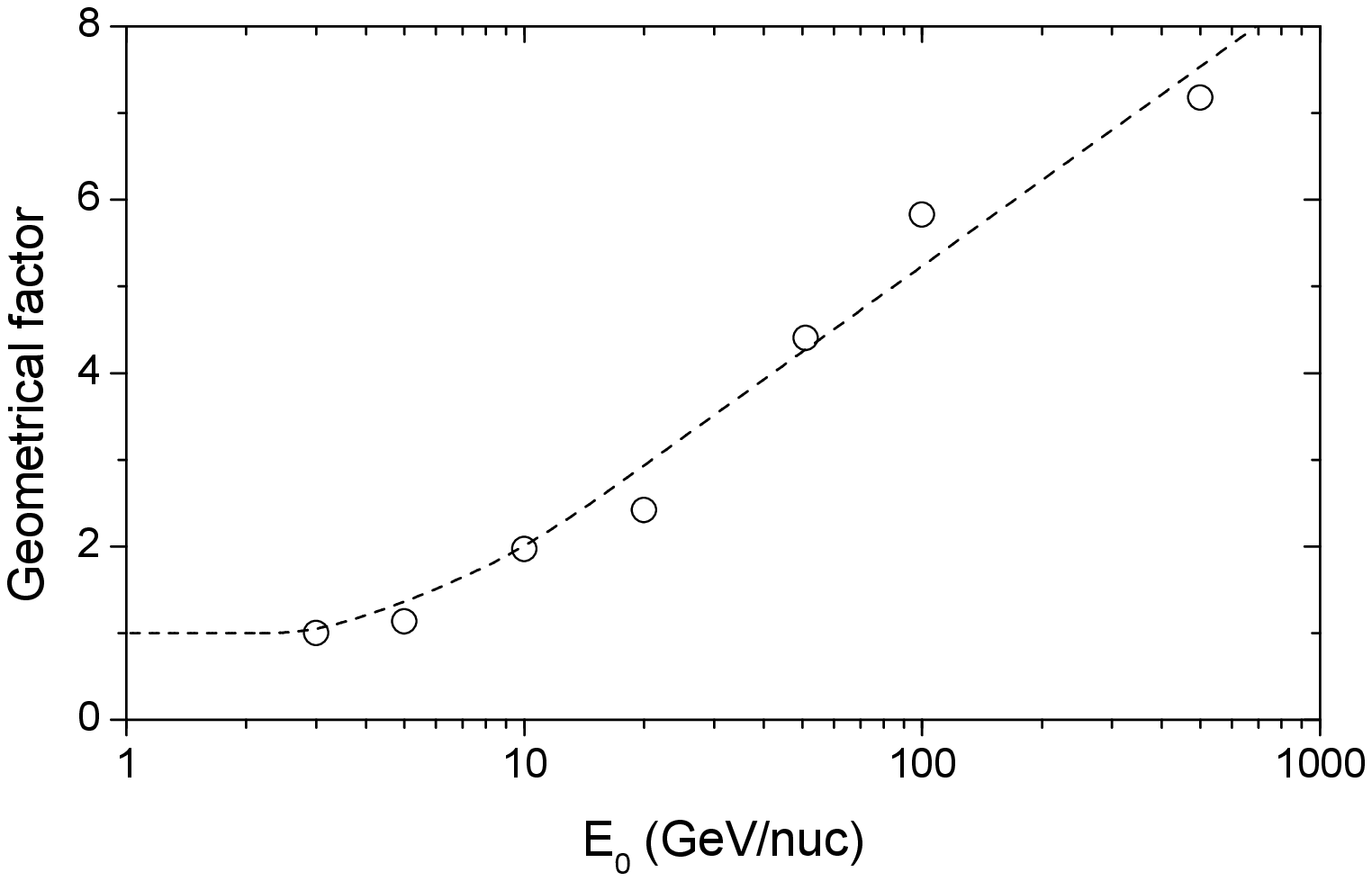}}
\end{center}
\caption{
The effective geometrical factor $G$ (Eq.~\ref{Eq:S_eff}) as function of the primary CR particle's energy for protons.
Dots are the results of the Monte-Carlo simulations and the dashed line is the used approximation.
\label{Fig:G}}
\end{figure}

We have performed the full Monte-Carlo simulation of this effect in the following way.
We simulated cascades, caused by a primary CR particle with rigidity $P$, whose axis lies
 at the distance $R$ and azimuth angle $\beta$ from the center of the NM (Fig.~\ref{Fig:S}),
 and calculated the number $N(R,\beta,P)$ of secondary hadrons with energy above 1 MeV hitting the detector (hatched area in the Figure).
This step was repeated $10^5$ time for each set of $R$ and $P$ with the uniform random azimuthal distribution
 so that it includes averaging over the angle $\beta$.
As an example, $N\approx 0.01$ is found for a 3 GeV proton, vertically impinging on the top of the
 atmosphere above the center of NM.
We note that this energy corresponds to the effective energy of a polar NM, viz. to the maximum of the integrand of Eq.~\ref{Eq:N}.
Then the weight of such a remote cascade was calculated as
\begin{equation}
w(R,P)=\min\left[{N(R,P)};\ 1\right].
\label{Eq:w}
\end{equation}
This weighting takes into account that a standard NM has a large dead-time (about 2 ms) which cuts off
 the multiplicity of the counts \citep{stoker00}.
In other words, if many secondary particles hit a NM counter, there is still only one count, and all
 the subsequent multiple counts are cut off by the dead-time of the detector.
Thus, we assume that, if several secondary particles from the same cascade hit the detector
 within the dead-time, the number of counts is not increased.
The considered effect will be greater for NMs with short dead-time to study multiplicities.
Then the effective geometrical factor of the NM with the account for this effect is calculated as
\begin{equation}
G(P) = {2\pi\over S_{\rm NM}}\int_0^\infty{R\cdot w(R,P)\cdot dR}
\label{Eq:S_eff}
\end{equation}
where $S_{\rm NM}$ is the geometrical area of a 6NM64.
The effective geometrical factor $G$ is shown in Fig.~\ref{Fig:G} as dots.
One can see that above the energy of 5 GeV it is a logarithmicaly growing function of energy.
For further calculations we used an approximation shown as the dashed line in the Figure.
We note that this geometrical factor is slightly greater for smaller NMs and smaller for bigger NMs.

Then the definition of the NM yield function becomes as the following (cf. Eq.~\ref{Eq:Y})
\begin{equation}
Y_{i}(P,h)=  G(P) \sum_{i} \int \int A_{i}(E,\theta)\cdot F_{i,j}(P,h,E,\theta)\ dE\ d\Omega
\label{Eq:Y_cor}
   \end{equation}

\subsection{The NM yield function}

\begin{figure}
\begin{center}
\resizebox{\columnwidth}{!}{\includegraphics{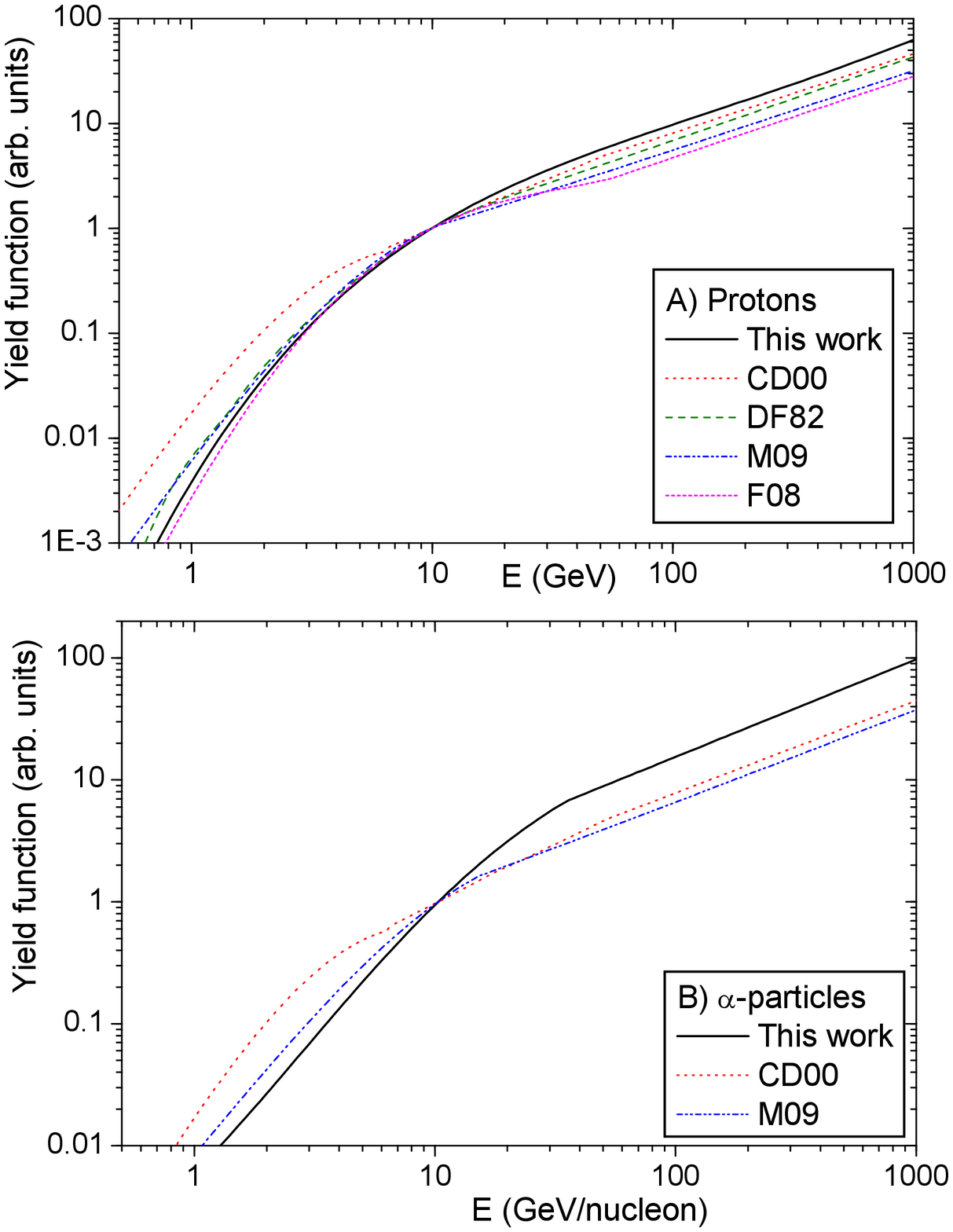}}
\end{center}
\caption{
The normalized (to unity at 10 GeV) yield function of the standard sea-level NM64 neutron monitor, calculated
 for CR protons (panel B) and $\alpha-$particles (panel B).
Different curves correspond to different models: CD00 \citep{clem99,clem00}; DF82 \citep{debrunner82};
 M09 \citep{matthiaPhD,matthia09}; F08 \citep{flueckiger08}.
When the high energy part (above 50--100 GeV) is not available from the original works, it has been
 extrapolated using a power law.
\label{Fig:YF_p}}
\end{figure}

Here we present the results for the new computations of the yield function of a standard neutron monitor,
 applying also the earlier neglected correction of the geometrical factor of a NM.
The tabulated values are given in Table~\ref{Tab:YF}.
Fig.~\ref{Fig:YF_p} presents the normalized yield functions of the standard sea-level 6NM64 neutron monitor
 computed for isotropically incident CR protons and $\alpha-$particles.

One can see that the yield function for protons (Fig.~\ref{Fig:YF_p}A) is close to earlier computations (except for that by \citet{clem00})
 in the lower energy range below about 10 GeV energy, but is greater than those in the higher energy range.
The latter is due to the geometrical correction factor (Fig.~\ref{Fig:G}).
The agreement in the lower energy range is particularly good with those \citep{flueckiger08,matthia09} based on the same GEANT-4 Monte-Carlo tool
 as our work.
The yield function for $\alpha-$particles has larger differences with respect to the earlier works, particularly \citep{clem00}
 who used FLUKA Monte-Carlo package.

\begin{table*}
\caption{Yield function (in units of [m$^2$ sr]) of the standard 6NM64 sea-level neutron monitor
 for the primary protons (columns 2--3) and $\alpha-$particles (columns 4--5), per nucleon of the primary particle.
\label{Tab:YF}}
\begin{tabular}{ccc|ccc}
\hline
\multicolumn{3}{c|}{protons} & \multicolumn{3}{c}{$\alpha-$particles}\\
$P$ (GV) & $E_{\rm p}$ (GeV) & Y$_p$ & $P$ (GV) & $E_\alpha$ (GeV/nuc) & Y$_\alpha$\\
\hline
0.7 & 0.232 & 7.26$\cdot10^{-9}$  &  &  & \\
1 & 0.433 & 8.46$\cdot 10^{-6}$ & 3.38 & 1 & 3.86$\cdot 10^{-4}$\\
2 & 1.27 & 1.21$\cdot 10^{-3}$ & 5.55 & 2 & 2.16$\cdot 10^{-3}$\\
3 & 2.21 & 5.42$\cdot 10^{-3}$ & 7.64 & 3 & 4.31$\cdot 10^{-3}$\\
4 & 3.17 & 9.43$\cdot 10^{-3}$ & 9.69 & 4 & 7.27$\cdot 10^{-3}$\\
5 & 4.15 & 0.02  & 11.7 & 5 & 1.18$\cdot 10^{-2}$\\
10 & 9.11 & 0.109  & 21.8 & 10 & 8.37$\cdot 10^{-2}$\\
20 & 19.1 & 0.229  & 41.8 & 20 & 0.215\\
50 & 49.1 & 0.59  & 100 & 49.1 & 0.59\\
100 & 99.1 & 0.992  & 200 & 99.1 & 0.992\\
500 & 499.1 & 3.35  & 1000 & 499.1 & 3.35\\
1000 & 999.1 & 6.67  & 2000 & 999.1 & 6.67\\
\hline
\end{tabular}
\end{table*}

\section{Verification of the model: Latitude survey}
\label{Sec:survey}

\begin{figure}
\begin{center}
\resizebox{\columnwidth}{!}{\includegraphics{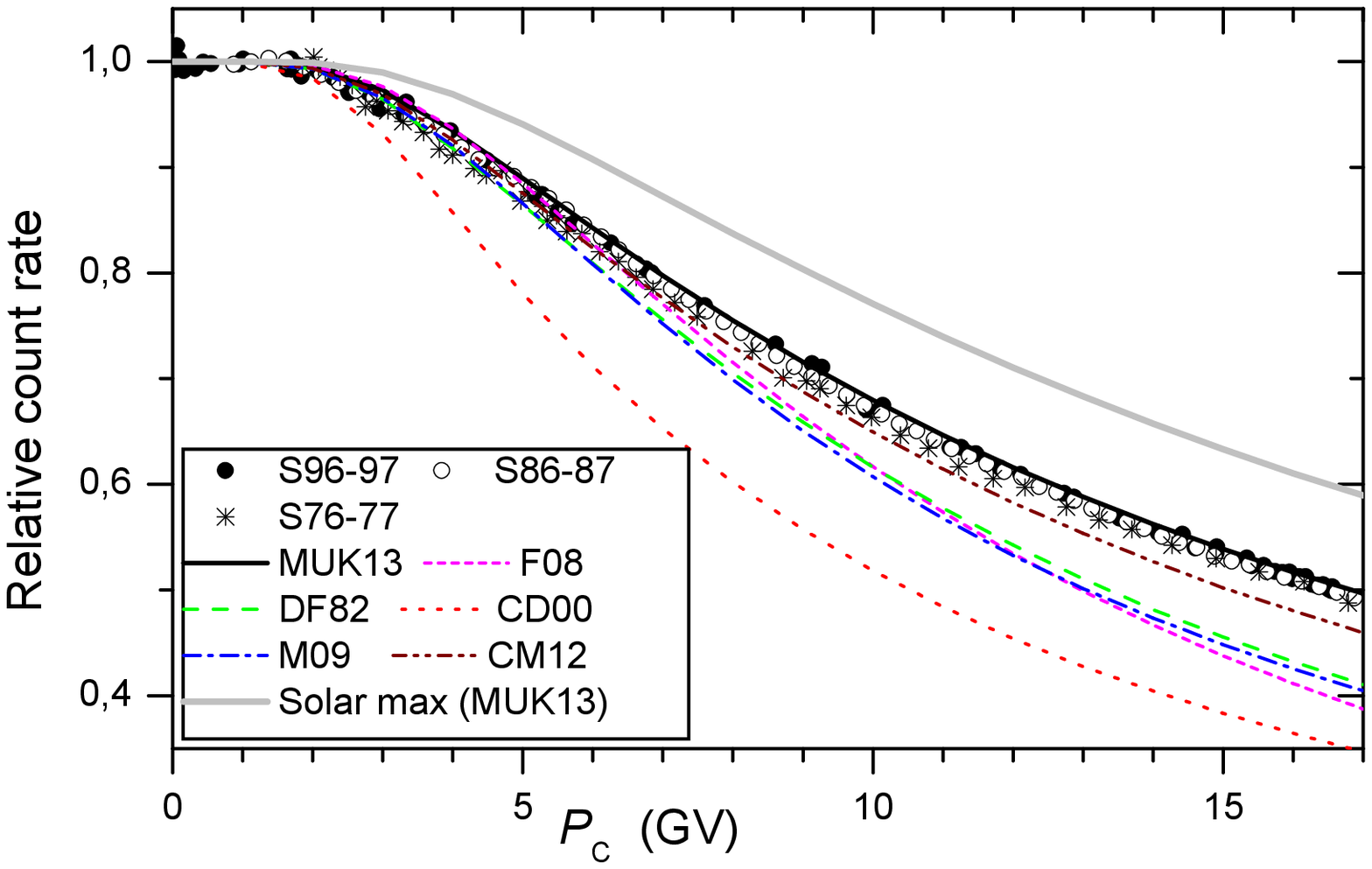}}
\end{center}
\caption{Latitude surveys.
Experimental data are depicted by symbols, while computed profiles are given as curves.
The data include surveys in 1996-97 \citep[][ - filled dots]{villoresi00}, 1986-87
 \citep[][ - open dots]{moraal89} and 1976-77 \citep[][ - asterisks]{potgieter79}.
The models are: this work [MUK13], \citep[][ - F08]{flueckiger08}, \citep[][ - DF82]{debrunner82},
 \citep[][ - CD00]{clem99, clem00}, \citep[][ - M09]{matthia09}, \citep[][ - CM12]{caballero12}.
All count rates are normalized to the polar region.
The grey curve depicts a survey calculated (using this model) for a solar maximum with $\phi=1200$ MV.
\label{Fig:survey}}
\end{figure}

As discussed above, a latitude survey, viz. change of the count rate of a standard NM surveying
 over latitudes, provides a direct test to the computation of the NM count rate and thus to the yield function.
Surveys are usually performed onboard a ship cruising between equatorial and polar regions.
Since such a cruise may take several months, they are made during the times of solar minimum in order to
 secure the least temporal variability of the cosmic ray flux so that all the changes in the surveying NM
 are caused by the changing geomagnetic shielding along the route \citep{moraal00}.
Here we consider three surveys, shown in Fig.~\ref{Fig:survey}, performed at consecutive solar
 minima in 1976--77 \citep{potgieter79}, 1986--87 \citep{moraal89} and 1996--97 \citep{villoresi00}.
The heliospheric conditions were similar during all the three surveys, with the modulation
 parameter being $\phi=400\pm 20$ MV \citep{usoskin_bazi_11}.
Accordingly, we will consider all the three surveys as one.
As the GCR spectrum we consider the force-field approximation with the fixed modulation parameter
 $\phi=400$ MV.

All the earlier theoretically calculated NM yield functions were unable to reproduce the observed
 latitude surveys \citep{clem00,caballero12}.
In order to test the yield functions, we have calculated the results of the latitude survey,
 predicted by these models, by applying Eq.~\ref{Eq:N} and using measured or
 modelled spectrum of cosmic rays \citep[see methodology in][]{usoskin_Phi_05}.
Here we consider the parametrization of the GCR energy spectrum \citep{usoskin_Phi_05}
 via the modulation potential $\phi$ and the local interstellar spectrum according to \citet{burger00}.
This parametrization is widely used in many applications and has been validated by comparisons with direct
 balloon- and space-borne GCR measurements \citep{usoskin_bazi_11,usoskin_ECRS_12}.
We note that there is an uncertainty in the exact shape of the GCR spectrum,
 and many other approximations exist \citep[e.g.,][and references therein]{herbst10}.
When applying other GCR spectral models, the results presented here would be slightly changed
 of the order of 10\%.
Computations were done separately for both protons and $\alpha-$particles, which effectively
 include also heavier CR species, similar to \citet{usoskin_bazi_11}.
Only CD00, M09 and the present work provide the NM yield function for primary $\alpha-$particles.
In all other cases we applied the proton yield function also for $\alpha-$particles.
The results of the computations are shown as curves in Fig.~\ref{Fig:survey}.
One can see that the lowest curve for the CD00 yield function overestimates the ratio between
 pole and equator by roughly a factor of two.
Other theoretical curves (DF82, F08 and M09) lie close to each other but about 20\% lower than the
 observed latitude survey.
The CM12 curve is based on an ad-hoc empirical yield function designed to fit the latitude survey.
Therefore, it is expected that it appears close to the measured profile.
A small 7--10\% discrepancy is caused by the different energy spectra of GCR used here and by \citet{caballero12}.
We note that, provided the same spectral shape is used, the CM12 result must lie on the
 observed latitude survey curve by definition.
The theoretical yield function presented in this work (MUK13) precisely reproduces the observed latitude survey,
 without any ad-hoc calibration or adjustment.
We note that our present yield function, but without the geometrical correction (Section~\ref{Sec:geom}),
 yields the results very similar to that by CD00.
Thus, the account for the geometrical correction fully resolves the problem of the NM yield function
 being in disagreement with the experimental latitude surveys of NMs.

\section{Conclusions}
\label{Sec:conc}

We have presented here a newly computed yield function of the standard sea-level neutron monitor 6NM64,
 separately for primary protons and $\alpha-$particles, the latter representing also heavier species of cosmic rays \citep{webber03}.
The computations have been done using the GEANT-4 tool Planetocosmics and a realistic atmospheric model.
For the first time, an important but previously neglected effect of the geometrical correction of the neutron monitor size, related to the
 finite lateral expansion of the CR-induced atmospheric cascade, is considered.
This correction enhances the relative impact of higher-energy cosmic rays (energy above 5--10 GeV/nucleon)
 in the neutron monitor count rate leading to weaker dependence of GCR on the geomagnetic shielding.
This improves the situation with the long-standing problem of disagreement between the theoretically calculated
 spatial variability of cosmic rays over the globe and the experimental latitude surveys.
The NM yield function, corrected for this geometrical factor, appears perfectly consistent with the
 experimental latitude surveys of neutron monitors performed during three consecutive solar minima
 in 1976--77, 1986--87 and 1996--97.
On the other hand, this geometrical correction does not affect analysis of events in solar energetic particles
 whose energy is usually below a few GeV.

Thus, we provide a new yield function of the standard sea-level neutron monitor 6NM64 that
 is validated versus confrontation with experimental data and can be applied in detailed studies of
 cosmic ray variability on different spatial and temporal scales.

\begin{acknowledgement}
This work was partly funded from the European Union's 7-th Framework Programme (FP7/2007-2013) under
   grant agreement N 262773 (SEPServer), by the Academy of Finland and by the V\"ais\"al\"a foundation.
We acknowledge the high-energy division of Institute for Nuclear Research and Nuclear Energy - Bulgarian Academy of Sciences for
 the computational time.
The authors warmly acknowledge John Clem and Boris Gvozdevsky for the
 updated information concerning the NM registration efficiency.
We acknowledge also Vladimir Makhmutov for information related to PLANETOCOSMICS simulations, as well as Eduard Vashenuyk and Yury Balabin
 for fruitful discussions concerning GLE analysis.
G.A.Kovaltsov was partly supported by the Program No. 22 of presidium of Russian Academy of Sciences.
\end{acknowledgement}




  %

\pagebreak

\end{article}
\end{document}